\documentstyle[prl,aps]{revtex}  
\begin{document}
\draft

\title{Phase coherent transport in hybrid superconducting structures:
 the case of d-wave superconductors } 

\author{P.M.A. Cook,  R. Raimondi, and  C.J. Lambert}
\address{
School of Physics and Chemistry,
Lancaster University, Lancaster LA1 4YB, U.K.}

\date{\today}
\maketitle

\begin{abstract}

We examine the effect of d-wave symmetry on
zero  bias anomalies in normal-superconducting tunnel
junctions and phase-periodic conductances in
Andreev interferometers.  In the
presence of d-wave pairing,  zero-bias anomalies
 are suppressed compared with the s-wave case.
For
Andreev interferometers with aligned islands,
the phase-periodic conductance is insensistive
to the  nature of the pairing, whereas for non-aligned islands, the
nature of  zero-phase extremum is reversed.
\end{abstract}
\section{Introduction}
\label{introduction}

During the past few years, studies of the subgap conductance
of  normal-insulator-superconductor (N-I-S)
 structures have revealed new and unexpected
behaviour\cite{ka91}. 
Whereas conventional tunneling theory suggests that the subgap
conductance must vanish,
experiments reveal that when the normal region
becomes  phase-coherent, there exists
a zero-voltage peak in the differential conductance,
which  can be comparable with the normal state value.
 The interplay between
Andreev scattering\cite{an64}
 at an N-S interface and  disorder-induced scattering in the
normal region  has been recognised 
as the main physical origin of this zero bias anomaly
 (ZBA) \cite{za90}-\cite{be94}.

Andreev scattering\cite{an64}
involves the simultaneous tunneling
of two electrons of
opposite momenta
 through the N-I-S structure.
In the case of a low-conductance tunnel junction
($G_{tun} \ll 1$),
 this two-particle process occurs with a
probability $\approx G_{tun}^2$.
 As a consequence, the
 conductance in the superconducting state
is expected to be much smaller than that in the normal state.
 However, disorder-induced  scattering  on the normal side
 may give   rise to particle-particle
correlations, which effectively, increase the
 probability for two particles with opposite
momenta  to tunnel into the superconductor.
This increase  manifests itself as a low
total-momentum singularity in the
particle-particle scattering channel (the so-called Cooperon)
\cite{ef80}, 
which is also the relevant scattering channel for
the occurrence of s-wave pairing  in BCS theory\cite{bcs}.
In the latter case, the low total-momentum instability signals the
formation of a bound state of two electrons with opposite
momenta, while
in the former  it leads to enhanced
backscattering in disordered systems.
In both cases, the singularity in the particle-particle Cooperon channel 
is characterized by the s-wave symmetry of the relative two-particle
wavefunction and  is ultimately responsible
for the occurrence of the ZBA\cite{he93}. In contrast,
for a clean normal region, Andreev scattering is almost
suppressed\cite{btk82} and conventional tunneling theory applies.

The aim of this paper is to demonstrate that  ZBAs and related
phenomena are  sensitive to the symmetry of the
superconducting order parameter and therefore can be
used to distinguish between s-wave and d-wave pairing
in high-$T_c$
superconductors. In what follows, we 
extend a numerical multiple scattering approach \cite{la94} to the case of
non-local anisotropic superconductors.
In the case of local s-wave
pairing, this approach has been shown to be equivalent
to other techniques
 and has been used to study the cross-over
between different transport regimes\cite{crl95}.
To start with,
in section II , we present an analytic treatment of a ballistic
N-I-S structure, which generalizes the theory of Ref.\cite{btk82} to the case of 
a tight-binding lattice and of a non-local superconducting order parameter.
In section III,  we show that, in the presence of disorder,
 the subgap conductance of a normal-insulating- d-wave junction 
  is {\it  suppressed}
compared with the corresponding normal-insulating-s-wave structure.
 In particular,  in the regime of small tunnel junction
 conductance $G_{tun}\ll 1$, we predict that the d-wave junction  shows only
 a weak  zero bias anomaly.
Similarly, the zero energy dip of a N-I-S structure,
in the high tunnel junction conductance
 regime $G_{tun} \approx 1$,
 is  considerably smaller for  the d-wave case.

Having examined the case of a single superconducting contact, in section IV, we
extend the analysis to a phase-coherent structure in contact with
two superconductors. The electrical conductance of such structures is known
to be a periodic function of the difference between the order parameters phases
of the two superconductors, which in turn is an
externally controllable quantity.
Such Andreev interferometers have recently
been the subject of intensive theoretical\cite{spi82}-\cite{na95}
and experimental studies
\cite{ve94}-\cite{pe95}. In what follows,
we predict that certain features of the phase-periodic conductance,
such as the nature of the zero-phase
extremum, are sensistive to the symmetry of the order parameter,
and, therefore, interferometers of this kind
provide a further probe into the nature of the pairing.

\section{Andreev reflection at a N-S interface}
\label{analytic}

To begin with, in this section,
 we examine a normal-superconducting
interface (N-S) in the presence of a non-local pairing potential,
described  by the  Hamiltonian\cite{scalapino95}

\begin{equation}
\label{an.1}
H=\sum_{i,\sigma}\epsilon_i c^{\dagger}_{i,\sigma}c_{i,\sigma} +
\sum_{i,\delta ,\sigma}\left[ \gamma c^{\dagger}_{i,\sigma}c_{i+\delta , \sigma}
+ h.c. \right]+
\sum_{i,\delta}\left[ (c^{\dagger}_{i,\uparrow}c^{\dagger}_{i+\delta,\downarrow}
-c^{\dagger}_{i,\downarrow}c^{\dagger}_{i+\delta,\uparrow}) \Delta_{i,\delta}
+h.c. \right].
\end{equation}
Here, the index $i$ runs over a two-dimensional
tight-binding lattice, with unit lattice constant,
 $\delta$ sums over  nearest neighbours, $\delta = \hat x, \hat y$, and  the pairing
potential, $\Delta_{i, \delta}$,  is defined on the  bond 
from site $i$ to site $i+\delta$. The operator $c_{i, \sigma}^{\dagger }$
($c_{i, \sigma}$ ) creates (destroys) an electron at site $i$, with 
  site energy $\epsilon_i$, and  $\gamma$ is the hopping matrix element
between nearest-neighbours sites. All energies will be measured in units
of $\gamma$, which will be set to unity throughout the paper.
The above Hamiltonian is diagonalised by solving 
 the Bogoliubov - de Gennes equation:

\begin{eqnarray}
\label{an.3}
\begin{array}{c c}
E\psi_i
=&\epsilon_i \psi_{i}
+\sum_{\delta} \gamma \left(  \psi_{i+\delta} + \psi_{i-\delta} \right)
+\sum_{\delta} \left( \Delta_{i,\delta} \phi_{i+\delta}
+\Delta_{i-\delta,\delta} \phi_{i-\delta}\right)\\
E\phi_i =&-
\epsilon_i \phi_{i}
-\sum_{\delta} \gamma  \left(  \phi_{i+\delta}+ \phi_{i-\delta}\right)
+\sum_{\delta} \left( \Delta^*_{i,\delta} \psi_{i+\delta}
+\Delta^*_{i-\delta,\delta}\psi_{i-\delta}\right),\\
\end{array}
\end{eqnarray}
where $\psi_i$($\phi_i$) indicates the particle (hole) wavefunction. 
In general, the Hamiltonian of eq.(\ref{an.1}) is a mean-field approximation
to
a more complex Hamiltonian containing electron-electron interactions and
all parameters should  be determined self-consistently.
However, in many cases of experimental
interest, the qualitative form of parameters such as
$\Delta_{i,\delta}$ is known and for the purpose of highlighting
generic transport properties, self-consistency is not required\cite{note}.

The Bogoliubov-de Gennes equation may be solved by means of  a
transfer matrix method or a recursive Green function technique. These
methods work in any dimension and constitute the only exact approach
when  translational invariance is absent.
As a prelude to such a calculation, in this section, we
 begin  by considering a system with
translational invariance in the direction perpendicular to the current flow.
Our motivation for doing so is two-fold. On the one hand, one obtains
new results for  Andreev scattering in the
presence of anisotropic pairing. On the other, the analysis
provides a controllable limit, which  can be  used to test  the numerical
 machinery
used in more complicated situations.

In the presence of translational invariance in the  direction
transverse to the current,
the problem reduces to one of 
 many independent
one-dimensional channels, each characterised by one or more
 discrete quantum numbers.
  In what follows, we  consider the case of a 2-dimensional system with a N-S
 interface, whose normal vector points in the $x$ direction, although
 by redefining the parameters $\bar\mu$, $\tilde\Delta$ introduced below,
 all results are trivially generalized to 3-dimensions.
The number of independent channels is determined by the width $M$
of the system,
and each channel has a discrete
 wavevector $k_y$, along the $y$ direction.
Choosing $\epsilon_i =-\mu$ yields an
energy dispersion relation in the normal region of the form

\begin{equation}
\label{an.5}
\epsilon_k =-2\gamma (cos(k_x) + cos(k_y))-\mu 
\end{equation}
where the uniform site energy $\mu$  determines the filling of the
 tight-binding band.
 Writing
\begin{equation}
\label{an.6}
\bar\mu=\mu+2\gamma cos(k_y).
\end{equation}
shows that the dispersion relation reduces to that of a one-dimensional system
with a channel-dependent chemical potential $\bar \mu$.

In the superconducting region, the dispersion relation (\ref{an.5}) is replaced by

\begin{equation}
\label{an.7}
E^2=(\epsilon_{\bar k})^2+|\Delta_{\bar k}|^2,
\end{equation}
where $\epsilon_{\bar k}=-2\gamma (cos(\bar k_x )+cos(\bar k_y))-\mu$, 
and the momentum dependent gap function, $\Delta_{\bar k}$, is given by

\begin{equation}
\label{an.8}
\Delta_{\bar k}=2 (\Delta_x cos(\bar k_x) +\Delta_y cos(\bar k_y)),
\end{equation}
where for d-wave symmetry,
 $\Delta_x =- \Delta_y $.

In writing the above equation, we have assumed a uniform  pairing potential 
for all sites $i>0$. The transverse component, $k_y$ is
 conserved through the interface, $\bar k_y = k_y$, and 
for a fixed value of  $k_y$, the gap function can be written

\begin{equation}
\label{an.10}
\Delta_{\bar k}=2 \Delta_x cos(\bar k_x) +\tilde\Delta ,
\end{equation}
where $\tilde\Delta =2\Delta_{y} cos(k_y)$, which demonstrates that
the pairing potential along the $y$ direction yields 
 a local contribution to the pairing potential for
a particular channel. Hence for $i< 0$, the Bogoliubov - de Gennes equation
reduces to

\begin{eqnarray}
\label{an.11}
\begin{array}{c c}
E\psi_i=& -\gamma(\psi_{i-1}+\psi_{i+1})-\bar\mu\psi_i \\
E\phi_i=&\gamma(\phi_{i-1}+\phi_{i+1})+\bar\mu\phi_i \\
\end{array}
\end{eqnarray}

while for $i>0$,

\begin{eqnarray}
\label{an.13}
\begin{array}{c c}
E\psi_i=&-\gamma(\psi_{i-1}+\psi_{i+1})-\bar\mu\psi_i
+\tilde\Delta \phi_i +\Delta_x (\phi_{i-1}+\phi_{i+1}) \\
E\phi_i=&\gamma(\phi_{i-1}+\phi_{i+1})+\bar\mu\phi_i +
\tilde\Delta\psi_i+\Delta_x (\psi_{i-1}+\psi_{i+1}).\\
\end{array}
\end{eqnarray}

Since
 the value of the transverse component of the momentum 
enters only through the effective chemical potential $\bar\mu$ and the 
local contribution to the gap $\tilde\Delta$,  it is  convenient
to drop the $x$ suffix in labelling the various longitudinal momenta and
write  $k$ and $q$, ($\bar k$ and $\bar q$), for the longitudinal
momenta of particles and holes in the normal (superconducting)
region. With this notation and writing
$(\psi^R, \phi^R )$ and  $(\psi^L, \phi^L )$ for the solutions when
$i>0$ and $i<0$, respectively, one obtains
 for $i<0$

\begin{eqnarray}
\label{an.15}
\begin{array}{c}
\psi_i^L =e^{ikR_i}+r_o e^{-ikR_i},\\
\phi_i^L =r_a e^{iqR_i},\\
\end{array}
\end{eqnarray}

and for $i>0$,

\begin{eqnarray}
\label{an.17}
\begin{array}{c c}
\psi_i^R =&t_o u e^{i\bar kR_i} + t_a \bar u e^{-i\bar q R_i},\\
\phi_i^R =& t_o v e^{i\bar k R_i} +t_a \bar v e^{-i\bar qR_i}. \\
\end{array}
\end{eqnarray}
 
 The coherence factors $u$ and $v$ identify a particle-like excitation of
energy $E$ and momentum $\bar k$, while $\bar u$ and $\bar v$ correspond
to an hole-like excitation at the same energy and momentum $\bar q$.
To compute these quantities one notes that

\begin{equation}
\label{an.27}
E^2=(-2\gamma cos(\bar p )-\bar\mu )^2 + |\tilde\Delta +2 \Delta_x cos(\bar p
)|^2
\end{equation}
where $\bar p$ may be $\bar k$ or $\bar q$, which yields

\begin{equation}
\label{an.28}
cos (\bar p )=-{1\over 2} \left( \bar\mu  + A \pm B\right)/\gamma
\end{equation}
with

$$
A=\Delta_x  {{\gamma \tilde\Delta -\Delta_x \bar\mu }\over
{\gamma^2 + \Delta_x^2}},~~~~~
B=\gamma \sqrt{ {{E^2}\over {\gamma^2 +\Delta_x^2}} -
\left( {{\gamma \tilde\Delta -\Delta_x \bar\mu }\over
{\gamma^2 + \Delta_x^2}}\right)^2}
$$

For the coherence factors one has to distinguish the case of real $B$,
which corresponds to quasi-particle transmission through the interface,
from the case of   imaginary $B$, which corresponds to  no
quasi-particle transmission.

For $B$ real, one obtains

\begin{eqnarray}
\label{an.29}
\begin{array}{c}
u^2 ( v^2 )={1\over 2} (1 +(-) (A+B)/ E)\\
{\bar u}^2 ( {\bar v}^2 )={1\over 2} (1 +(-) (A-B)/ E)\\
\end{array}
\end{eqnarray}

with $u/v=sign \left({ \Delta_{\bar k}/(E-\epsilon_{\bar k})}\right)$, and
 ${\bar u}/{\bar v}=sign\left( { \Delta_{\bar q}/(E-\epsilon_{\bar
q})}\right)$.

For $B$ imaginary, we write  $B\equiv i \tilde B$, to yield

\begin{eqnarray}
\label{an.31}
\begin{array}{c}
u/v=\Delta_{\bar k} / (E-A-i\tilde B)\\
{\bar u}/{\bar v}=\Delta_{\bar q} / (E-A+i\tilde B)=(u/v)^*.\\
\end{array}
\end{eqnarray}

To solve for the scattering coefficients $r_o$, $r_a$, $t_o$, and $t_a$,
one needs matching conditions at the interface.
These are obtained by evaluating equations (\ref{an.11}) 
and (\ref{an.13}) at $i=0$ and
at $i=-1$ and can be written in the form

\begin{eqnarray}
\label{an.19}
\begin{array}{c}
\psi_0^L=\psi_0^R\\
\phi_0^L=\phi_0^R\\
\end{array}
\end{eqnarray}

and

\begin{eqnarray}
\label{an.21}
\begin{array}{c}
\psi_{-1}^L=\bar\psi_{-1}^R\\
\phi_{-1}^L=\bar\phi_{-1}^R,\\
\end{array}
\end{eqnarray}
where (see appendix) $\bar\phi_{-1}^R$, $\bar\psi_{-1}^R$ are obtained by acting on
$\phi_{i}^R$, $\psi_{i}^R$ with the appropriate transfer matrix  at  the
interface.

Equation (\ref{an.19})  yields

\begin{eqnarray}
\label{an.23}
\begin{array}{c} 
1+r_o=t_o u  + t_a \bar u\\
r_a=t_o v  + t_a \bar v,\\
\end{array}
\end{eqnarray} 
whereas the matching conditions eq.(\ref{an.21})
yield (see appendix)

\begin{eqnarray}
\label{an.25}
\begin{array}{c}
e^{-ik} +r_o e^{ik}=
t_o u e^{-i\bar k} (1- (v/u)(\Delta_x /\gamma ))
+t_a \bar u e^{i\bar q} (1-(\bar v /\bar u ) (\Delta_x / \gamma ))\\
r_a e^{-iq}=
t_o v e^{-i\bar k} (1+ (u/v)(\Delta_x /\gamma ))
+t_a \bar v e^{i\bar q} (1+(\bar u /\bar v ) (\Delta_x / \gamma )).\\
\end{array}
\end{eqnarray}

The  eqs.(\ref{an.23}-\ref{an.25}) yield
all the scattering coefficients associated with a particle incident from
 the left on a clean  N-S interface.
A direct analytic evaluation is rather messy, but it is trivial to solve
these  numerically. Explicit results will be presented in  section III. 
In the presence of a tunnel barrier modelled by
 a delta function potential at the
interface, $\epsilon_i$ is replaced by
 $\epsilon_i =-\mu +U\delta_{i,0}$ and the
matching conditions eq.(\ref{an.25})
by

\begin{eqnarray}
\label{an.33}
\begin{array}{c}
e^{-ik} +r_o e^{ik}=
t_o u e^{-i\bar k} (1- (v/u)(\Delta_x /\gamma ))
+t_a \bar u e^{i\bar q} (1-(\bar v /\bar u ) (\Delta_x / \gamma ))
+(1+r_o )U/\gamma\\
r_a e^{-iq}=
t_o v e^{-i\bar k} (1+ (u/v)(\Delta_x /\gamma ))
+t_a \bar v e^{i\bar q} (1+(\bar u /\bar v ) (\Delta_x / \gamma ))
+ r_a U/\gamma.\\
\end{array}
\end{eqnarray}

Various limiting forms of the above expressions are discussed  in appendix.
Here we merely note that  with the convention adopted in
eqs.(\ref{an.15}-\ref{an.17}),
if $v_k$ ($v_q$) and $v_{\bar k}$ ($v_{\bar q}$)
are group velocities for particles (holes) in the normal and superconducting
regions respectively, then the following unitarity condition is satisfied,
$R_a+R_0+T_a+T_0=1$, where $R_a=(v_q/v_k)\vert r_a\vert^2$,
$R_0=\vert r_0\vert^2$, $T_a=(v_{\bar q}/v_k)\vert t_a\vert^2$
and $T_0=(v_{\bar k}/v_k)\vert t_0\vert^2$.

\section{Results for N-I-S structures}

In this section we present explicit results for the above scattering
coefficients and for the electrical conductance\cite{la91}

\begin{equation}
G=\left( {{2e^2}\over{h}}\right) \sum_{k_y} 
\left( 1-R_0 +R_a\right)
\label{3.1}
\end{equation}
where the sum is over all channels.

For an  N-S interface
with no potential barrier, fig.\ref{fig1}(a)  shows
the behaviour  of the differential conductance as a function of energy, 
along with
various scattering coefficients. 
In a conventional superconductor under sub-gap conditions,
only Andreev reflection contributes to the current flow, because
sub-gap quasi-particle  transmission is forbidden.
In a gapless superconductor, the situation is more complicated,
because each channel
has its own effective gap, so that normal quasi-particle
transmission
occurs even at low energies. To illustrate this point,
 fig.\ref{fig1}(a) shows the behaviour of  transmission
and reflection probabilities
 for a system width $M=10$. 
The number of open channels $N$ depends on the position of the chemical
potential $\mu$
within the band. Here we have used $\mu =-0.2\gamma$ so that $N=9$.
For free-end
 boundary conditions in the transverse direction, the allowed values
of $k_y$ are  $\pi n/(M+1)$, with $n=1,2 ...M$.
For this  choice, all channels have non-vanishing gap and
at zero energy,  normal transmission vanishes.
 By increasing the energy, one eventually  crosses the effective gap of
a particular channel, at which point transmitting 
channels appear in the superconducting region
 and  Andreev scattering is suppressed. Since
Andreev scattering contributes  a factor of $2$ 
(in units of $2e^2/h$) to the electrical
conductance, while transmission processes only contribute a factor
of unity,  the conductance decreases at such energies.
In fig.\ref{fig1}(b), we  show  results for a system width
 $M=1000$. At
zero energy, it is  interesting to note that the conductance per
 channel is insensitive to the width of the system and
 agrees perfectly with the
 $M=10$ value.
At finite energy the results  differ slightly,
because in the case of a large number of channels,
channels open  continously with  increasing energy, to yield the smooth
behaviour shown in fig.\ref{fig1}(b).

We now consider 
the case of an N-I-S structure, in which a tunnel junction is 
present at the interface.
Fig.\ref{fig2}(a-b) shows results for the electrical conductance
$G$ and for various scattering coefficients  in the presence of a
 barrier height  $U=2.0\gamma$ for a system width $M=10$.
The oscillating behavior is  a finite size effect which can be understood by 
 observing that a peak occurs when the energy becomes greater 
than the effective gap $\tilde\Delta$ of a particular channel.
The  oscillation arises because for $E < \tilde\Delta$ the contribution
from such a channel is proportional to the square of the barrier transmission
coefficient $\Gamma$ (i.e., $\approx G_{tun}^2$), 
whereas for $E > \tilde\Delta$ it is proportional
to the first power of $\Gamma$ (i.e., $\approx G_{tun}$).

In fig.\ref{fig2}(c-d) we show the conductance for a width $M=1000$.
Again the conductance per channel, at zero energy,  is found to be insensitive
to the number of open channels, and oscillations 
at intermediate energies are no-longer present.

Having examined a clean  metal in contact  with a superconductor,
we now introduce disorder to the normal metal, to produce a diffusive
conductor in contact with a d-wave superconductor. In this case
the conductance is obtained using the transfer matrix method outlined in
reference \cite{lhr93}  and computer resources impose restrictions on
 the system width $M$.

The N-I-S structure of fig.\ref{fig3}  consists
of a diffusive metallic region  placed in series  with a tunnel junction,
 which in turn is adjacent to a superconductor.   In the absence of
 disorder, the numerical code agrees exactly with the analytical
 results of figures \ref{fig1} and \ref{fig2}. In what follows,
the simulated structure is a two-dimensional tight-binding lattice of width
 $M=10$ sites. The disordered region is of length $L_{dif}$ sites,
 the tunnel junction is $L_{tun}$ sites
 long and the superconductor has a length
$L_{sup}$. The conductance of the entire structure is denoted by $G$ and the
average over an ensemble of disorder realizations is $<G>$.
The physical variables in the following calculation
are the averaged conductance  $<G_{dif}>$ of the diffusive region and the 
conductance of the tunnel junction $G_{tun}$. To identify a suitable choice
 of parameters, we considered first a normal diffusive portion of length $L_{dif}$
 and width $M$, connected to crystalline, normal
 leads. The conductance of a diffusive
 material is inversely proportional to its length and therefore a plot
of $<G_{dif}>L_{dif}$ as a function of $L_{dif}$ will exhibit a plateau
in the diffusive regime, with a mean free path given by 
$l=<G_{dif}>L_{dif}/M (2e^2/h)$. A diffusive system must satisfy
$l\ll L_{dif}$ and $l\ll M$. Furthermore, if weak localization corrections  are
to be neglected, we require $Nl\gg L_{dif}$. In the calculations
which follow, having in mind also
the necessity of minimizing the CPU time, we have made the following choice of
 parameters: $L_{dif}=30$ sites, disorder width $W=1$, system width
$M=10$, $\mu =0$ so that the number of open channels is  $N=10$.
 This yields a mean free path $l=4.2$ and an average conductance 
$<G_{dif}>=1.6$ (in units of $2e^2/h$).
 The superconductor has a length $L_{sup}=100$ with order parameter
$\Delta_x =0.1\gamma $.
One characteristic energy scale is the maximum
gap, which  for $\Delta_x =0.1\gamma$ is equal
to $0.4\gamma$.
The tunnel junction is $L_{tun}=1$ site long and the potential
on the line of sites defining the junction is
$-\mu+\epsilon_b$ 
with $\epsilon_b$ taking the value  $7\gamma$ and $2\gamma$, which yields
$G_{tun}=0.4(2e^2/h)$ and $G_{tun}=3(2e^2/h)$ in the low and high tunnel
junction regimes, respectively.

For the case of low tunnel junction
 conductance, fig.(\ref{fig4}) shows
 the sub-gap { conductance} as a function  of the energy, which in the linear
 response regime, corresponds to
experimentally measured I-V characteristic.
The strong peak, present in the s-wave case, is strongly
suppressed in the d-wave case.
Fig.(\ref{fig5}) shows the corresponding behaviour in
the high tunnel junction conductance  regime.
In this case,  the zero energy behaviour is characterized 
by a dip in the conductance. As before,  the d-wave conductance and the zero
bias feature is
suppressed.

\section{Results for d-wave Andreev interferometers.}

 We now consider the effect of d-wave symmetry on the properties of
Andreev interferometers.
Two  different interferometer geometries are analyzed below and shown as
inserts in figures 6 and 7.
In each example, the system consists of two  superconducting
regions  with order parameter phases $\phi_1$ and $\phi_2$,
separated by  a normal region $N$, with a quasi-particle
current flowing vertically.
In figure 6, the S-N-S
structure is  placed in contact with
 a tunnel junction and  the
N and S regions are clean.
In figure 7, there is no  tunnel junction, but
 the whole S-N-S structure is disordered.

The physical parameters used to obtain figure  6
are as follows:  $\mu=0$ and the  tunnel barrier site-energy
$\epsilon_b=2\gamma$, which corresponds to
 an average conductance per channel of
 $G_{tun}/N=0.176 (2e^2/h)$.
As shown in figure 6, the electrical conductance in the presence of
 s-wave pairing shows a large
amplitude of oscillation and a zero-phase minimum, which are
characteristic
of a ballistic structure, as discussed in
Refs.\cite{crl95},\cite{asrl95}.  In contrast, for a d-wave interferometer,
 the nature of the zero phase extremum
 depends on the relative orientation of the two superconducting
islands. In one
case (solid line in figure \ref{fig6}) the x axes with a positive value 
of
the order parameter are parallel in both islands, while in the other case 
(dashed line in figure \ref{fig6}) the x and y axes are exchanged in
going
 from one island to the other.
 For the case of parallel orientation, the d-wave
interferometer is almost identical to  the s-wave case, exhibiting a large
 amplitude
 of oscillation and a zero-phase minimum. When the islands are orientated
perpendicular  to one another, the minimum becomes a maximum and the
 entire
 curve is
shifted by $\pi $.  This arises because
 an electron 
Andreev reflected at an N-S interface, aquires
 the phase of the bonds in the longitudinal direction 
(with respect to the  incoming direction of the electron).
When the two islands are orientated parallel,  they have the same value for
 the
longitudinal bonds, so that there is no effective phase difference between 
the 
two islands, as in the s-wave case. When one of the two superconducting islands
 is rotated by $\pi /2$,  the longitudinal bonds
have a phase difference of $\pi$, which adds to the external phase 
difference.
This effect is similar in origin to that predicted \cite{sr92}
and later  verified \cite{wo93} in a corner SQUID experiment on the
high-$T_c$ superconducting compound YBCO. 
The relevance
of the orientation of the crystal axis with respect to the N-S interface
 has
been also recently put forward\cite{yu95} as a possible explanation of
zero bias anomalies.

For figure 7  we   choose parameters corresponding to
 a diffusive region.
The structure has a length $L=50$ with each of the three regions having
a width $M'=15$, so that the entire structure has a width $M=45$. We
used $\mu=0$, disorder width $W=0.5\gamma$, which yields, in the normal state,
a mean free path $l=12$. 
For an s-wave
  interferometer (shown as an inset), the zero phase extremum switches
to a maximum and
the value at $\phi =\pi$ becomes a minimum.
For aligned islands (solid line),
the d-wave interferometer shows
the same qualitative behaviour as in the the s-wave case.
However, 
rotating one of the islands by $\pi /2$ again shifts the
curve by $\pi$ and changes the zero-phase
 extremum
from a maximum to a minimum.
\section{Discussion}

Zero  bias anomalies and phase-periodic conductances in
Andreev interferometers are paradigms of phase-coherent transport in
hybrid N-S structures. In this paper we have examined the sensistivity
of these phenomena to the symmetry of the superconducting order
 parameter. We find that ZBAs are suppressed,
reflecting the fact that disorder induced scattering and Cooper pairing
no longer occur in the same  particle-particle scattering channel.
We also find that for aligned islands,
Andreev interferometers are relatively insensistive
to the  nature of the pairing. Neverthless, such devices could be used
to reveal the presence of d-wave  pairing, since the positions of
conductance extrema are sensitive to the relative orientation of
the two order parameters.

For simplicity, we have avoided the necessity of a fully self-consistent
calculation, by restricting the analysis to a particular crystal orientation.
Analyses\cite{note}  of the proximity effect in ballistic systems reveal that a
self-consistent theory can yield a drastic suppression of the
superconducting order parameter in the vicinity of the N-I-S interface, but 
only for certain crystal orientations of the d-wave superconducting order
parameter with respect to the interface\cite{note2}.
In this paper, our main focus has been the interplay of Andreev scattering at
the interface and disorder-induced scattering in the normal region, and
therefore we have not considered these particular crystal orientations.
For this reason we expect that our main conclusions will not be qualitatively
changed by a fully self-consistent calculation, but for the future it would be
of interest to explore transport in the presence of alternative crystal
orientations.

\acknowledgements
We would like to thank   A.D. Zaikin 
for extended discussions. Financial support from the EPSRC, the MOD,
the Institute for Scientific Interchange, the E.C. H.C.M. program,  and
NATO is also gratefully acknowledged.

\appendix
\section{Technical details  of the analysis of section  II}
\label{a}
To obtain equation
(\ref{an.25}) one notes that the left-hand sides of
 eq.(\ref{an.21})
are

$$
\begin{array}{c}
\psi_{-1}^L=-(({{E+\bar\mu})/{ \gamma}}) \psi_{-2}^L -\psi_{-3}^L
=e^{-ik} +r_o e^{ik}\\
\phi_{-1}^L = (({{E-\bar\mu })/{\gamma}} )\phi_{-2}^L - \phi_{-3}^L
= r_a e^{-iq},\\ \nonumber
\end{array}
$$
  
whereas the right-hand sides are

$$
\begin{array}{c}
\gamma \bar\psi_{-1}^R = -(E +\bar\mu )\psi_0^R - \gamma \psi_1^R
+ \tilde\Delta \phi_0^R + \Delta_x \phi_1^R\\
\gamma \bar\phi_{-1}^R =(E-\bar\mu )\phi_0^R- \gamma \phi_1^R
-\tilde\Delta \psi_0^R -\Delta_x\psi_1^R.\\ \nonumber
\end{array}
$$

Notice that  $\bar\psi_{-1}^R$ and $\bar\phi_{-1}^R$ are decoupled because there is no
superconducting bond between $i=-1$ and $i=0$.
Inserting eq.(\ref{an.17}) into the above expressions
 and  using the eigenvalue equations, yields

$$
\begin{array}{c}
\bar\psi_{-1}^R =t_o u e^{-i\bar k} (1- (v/u)(\Delta_x /\gamma ))
+t_a \bar u e^{i\bar q} (1-(\bar v /\bar u ) (\Delta_x / \gamma ))\\
\bar\phi_{-1}^R =t_o v e^{-i\bar k} (1+ (u/v)(\Delta_x /\gamma ))
+t_a \bar v e^{i\bar q} (1+(\bar u /\bar v ) (\Delta_x / \gamma ))\\ \nonumber
\end{array}
$$ 

Finally one  obtains the last two matching conditions in the form
(\ref{an.25}).

The equations  obtained in section II are very general and
yield a variety of useful results. To illustrate this we
first rewrite
 equation (\ref{an.33})
in  the form

\begin{eqnarray}
\label{an.36}
\displaystyle{
\begin{array}{c}
t_o u {{e^{ik}- e^{-i\bar k} (1- (v/u)(\Delta_x /\gamma ))-(U/\gamma )}
\over{e^{ik}-e^{-ik}} }
+t_a \bar u {{e^{ik}-e^{i\bar q} (1-(\bar v /\bar u ) (\Delta_x / \gamma
))-(U/\gamma)}
\over{e^{ik}-e^{-ik}  }}=1\\
t_o v {{e^{-iq}- e^{-i\bar k} (1+ (u/v)(\Delta_x /\gamma ))-(U/\gamma )}
\over{e^{ik}-e^{-ik}} }
+t_a \bar v {{e^{-iq}-e^{i\bar q} (1+(\bar u /\bar v ) (\Delta_x / \gamma
))-(U/\gamma)}
\over{e^{ik}-e^{-ik}  }}=0.\\
\end{array}}
\end{eqnarray}

Once the above pair of equations is solved for $t_o$ and $t_a$,
eq.(\ref{an.23}) yields $r_o$ and $r_a$.
In the case of no quasi-particle transmission,
the only physical quantities  are $r_o$ and $r_a$, which, from
the form of eq.(\ref{an.36}),  depend only
upon the ratios $u/v$ and $\bar u /\bar v$. 

Results corresponding to the case of a local order parameter $\Delta_o$
 can be obtained from the above
eqs.(\ref{an.36}) by simply setting $\Delta_x =0$, and $\tilde\Delta
=\Delta_o$. By considering for simplicity an ideal interface ($U=0$), one
obtains

\begin{eqnarray}
\label{c1}
\displaystyle{
\begin{array}{c}
t_o u {{e^{ik}- e^{-i\bar k}}
\over{e^{ik}-e^{-ik}} }
+t_a \bar u {{e^{ik}-e^{i\bar q} }
\over{e^{ik}-e^{-ik}  }}=1\\
t_o v {{e^{-iq}- e^{-i\bar k} }
\over{e^{ik}-e^{-ik}} }
+t_a \bar v {{e^{-iq}-e^{i\bar q}}
\over{e^{ik}-e^{-ik}  }}=0.\\
\end{array}}
\end{eqnarray}

which gives

\begin{equation}
\label{c3}
t_o=\bar v (e^{i\bar q} - e^{-i q} ) ( e^{ik} -e^{-ik})/d
\end{equation}

\begin{equation}
\label{c4}
t_a= v (e^{-i q} - e^{-i \bar k} ) ( e^{ik} -e^{-ik})/d
\end{equation}

and, by using eq.(\ref{an.23}), 

\begin{equation}
\label{c5}
r_o=\left[ \bar v u (e^{i\bar q} -e^{-iq})(e^{-i\bar k} -e^{-ik})
+\bar u v (e^{-iq} -e^{-i\bar k})(e^{i\bar q} -e^{-i k})\right] /d
\end{equation}

\begin{equation}
\label{c6}
r_a= \bar v v (e^{i\bar q} - e^{-i \bar k} ) ( e^{ik} -e^{-ik})/d
\end{equation}

where 

$$
d=\left[ \bar v u (e^{i k} -e^{-i\bar k})(e^{i\bar q} -e^{-iq})
-\bar u v (e^{ik} -e^{i\bar q})(e^{-i\bar k} -e^{-iq})\right].
$$

Eqs.(\ref{c3}-\ref{c6}) solve the problem of determining the various
scattering coefficients. As a last step one has to substitute the 
expressions for the momenta $k$, $q$, $\bar k$, and $\bar q$ in terms 
of the energy $E$ as given by eq.(\ref{an.28}). We will not do this
substitution here, because the resulting equations
look rather cumbersome. We will instead content ourselves by considering the
relevant limit of zero quasiparticle energy. In this case there is no
quasiparticle transmission through the interface, and the momenta
$\bar k$ and $\bar q$ are complex. By setting
$
\bar k (\bar q) =p +(-)il
$
and observing that $q=k$ and 
$
\bar \mu =-2cos(k)=-2cos (p)cosh(l),
$
$
\Delta_0 = 2sin (p) sinh (l)
$
yields for the Andreev reflection scattering probability

\begin{equation}
\label{c7}
R_a=(v_q /v_q)|r_a |^2=
{{2(1- (\bar \mu^2 /4\gamma^2 ))}\over 
{ 1- (\bar \mu^2 /4\gamma^2 ) +(\Delta_o^2/4\gamma^2)+
\sqrt{(1+(\bar \mu^2 +\Delta_o^2)/4\gamma^2))^2-\bar \mu^2}}}.
\end{equation}

It is clear from  the above equation that in the limit 
$\Delta_o \rightarrow 0$, the Andreev scattering probability
$R_a \rightarrow 1$, which amounts to the so-called
 Andreev's  approximation largely used in a number of theoretical
treatments. To close this appendix, 
it is useful to make connection with the continuum
limit, which is most conveniently obtained from
 eqs.(\ref{c3}-\ref{c6}).  To make this
 connection,
we restore the lattice step $a$ in all expressions
involving the momenta, i.e., $k \rightarrow ka$ and take the limit
$a \rightarrow 0$,
$\gamma \rightarrow \infty$, such that 
$\gamma a^2 =\hbar^2 /2$. This leads to
\begin{equation}
r_o = \left[
u\bar v (q + \bar q )(k -\bar k) -
\bar u v (q-\bar k) (\bar q + k )
\right] / d
\label{a.5}
\end{equation}

\begin{equation}
r_a = \left[
2 v\bar v k (\bar k + \bar q ) 
\right] / d
\label{a.6}
\end{equation}

\begin{equation}
t_o = \left[
2 \bar v k (q + \bar q )
\right] / d
\label{a.7}
\end{equation}

\begin{equation}
t_a = \left[
-2  v k (q - \bar k )
\right] / d
\label{a.8}
\end{equation}

where 

$$d=u\bar v (k+\bar k ) (q + \bar q )-
\bar u v (q -\bar k ) (k -\bar q ).$$

In the limit ($\Delta_o /\mu \ll 1$),
 Andreev's approximation of ignoring both Andreev transmission
and normal reflection is valid and one obtains\cite{btk82}

\begin{equation}
R_a=1,~~~~E< \Delta_o
\label{a.9}
\end{equation}
and
\begin{equation}
R_a={{E-\sqrt{E^2-\Delta_o^2}}
\over{E+\sqrt{E^2-\Delta_o^2}}},~~~~E> \Delta_o
\label{a.10}
\end{equation}

In contrast, at zero energy, without invoking Andreev's approximation,
one obtains the exact result
\begin{equation}
R_a={{2\mu}\over{\mu +\sqrt{\mu^2+\Delta_o^2}}}
\label{a.11}
\end{equation}

with $R_o=1-R_a$. The above equation clearly shows how 
Andreev's  approximation breaks-down in the large $\Delta_o$ regime.

\begin{figure}
\caption{Electrical conductance (thick solid line),
Andreev reflection (solid line line), normal transmission (dashed
line).
Normal reflection and Andreev transmission are negligible on this scale.
  Parameters used are:  (a) width $M=10$, number of open channels $N=9$;
(b) width $M=1000$, number of open channels $N=857$; in all cases
$\mu =-0.2\gamma$, $\Delta_x =0.01\gamma$, barrier height $U=0$.}
\label{fig1}
\end{figure}

\begin{figure}
\caption{Electrical conductance (thick solid line),
Andreev reflection (solid line), normal transmission  (dashed line),
figures (a) and (c);
 normal reflection  (solid line),
 Andreev transmission  (dashed line), figures (b) and (d). 
  Parameters used:  (a) and (b) width $M=10$, number of open channels $N=9$;
(c) and (d) width $M=1000$, number of open channels $N=857$; in all cases
$\mu =-0.2\gamma$, $\Delta_x =0.01\gamma$, barrier height $U=2\gamma$.
Values are normalised to
the normal state conductance in order to compare with the case of no potential
barrier at the interface.}
\label{fig2}
\end{figure}

\begin{figure}
\caption{Schematic picture of an N-I-S stucture, considered in the text.
 Parameters used are: width $M=10$, lenght of the diffusive region $L_{diff}=30$, 
 length of the tunnel junction $L_{tun}=1$, disorder width $W=1$, 
$\mu =0$, number of open channels $N=10$.}
\label{fig3}
\end{figure}

\begin{figure}
\caption{Total conductance  as function of energy in the low tunnel 
junction
conductance regime. s-wave: dotted line; d-wave: solid line. $G_{tun}=0.4$
 in units of $2e^2/h$. The conductance is normalised to the normal state
conductance $G_o$. The two insets (left d-wave and right s-wave)
show the conductance behaviour on a more
extended energy range. The value of the superconducting order parameter in the
s-wave case is $\Delta_s =0.1\gamma$.} 
\label{fig4} 
\end{figure}

\begin{figure}
\caption{Total conductance as function of the energy in high tunnel
 junction conductance
regime. S-wave: dotted line; d-wave: solid line. $G_{tun}=3$ in units of $2e^2/h$.
The conductance is normalised to the normal state
conductance $G_o$.  The two insets (left d-wave and right s-wave)
show the conductance behaviour on a more
extended energy range.  The value of the superconducting order parameter in the
s-wave case is $\Delta_s =0.1\gamma$. }
\label{fig5} 
\end{figure}

\begin{figure}
\caption{The figure shows
results for a system  with an insulating barrier placed at one
 end
of the interferometer (See the structure displayed in the top
right corner).  The solid line (dashed line) indicates results 
for the conductance with aligned (not aligned) superconducting islands, 
for
a d-wave interferometer. The inset shows the corresponding result
for an s-wave interferometer.}
\label{fig6} 
\end{figure}

\begin{figure}
\caption{The figure shows the conductance  for a system  
without a barrier but when the 
interferomter is dirty (See the structure displayed in the top right
corner).  The solid line (dashed line) indicates results 
for the conductance with aligned (not aligned) superconducting islands, 
for
a d-wave interferometer. The inset shows the corresponding result
for an s-wave interferometer. }
\label{fig7}
\end{figure}


\begin{references}
\bibitem{ka91} A. Kastalsky, A.W. Kleinsasser, L.H. Greene,
F.P. Milliken, and J.P. Harbison, Phys. Rev. Lett. {\bf 67}, 3026 (1991).
\bibitem{an64} A.F.Andreev, Zh.Eksp.Teor.Fiz. {\bf {46}},1823 (1964);
[Sov.Phys.-JETP {\bf {19}}, 1228(1964)].
\bibitem{za90} A.V.Zaitsev, Sov. Phys. JETP Lett.  {\bf 51}, 41 (1990).
\bibitem{vo92} A.F. Volkov and T.M.Klapwijk,
 Phys. Lett. A {\bf 168}, 217 (1992).
\bibitem{vo93}  A.F. Volkov, Phys. Lett. A {\bf 174}, 144 (1992).
\bibitem{vzk93} A.F. Volkov, A.V. Zaitzev, and T.M. Klapwijk 
Physica C{\bf 210}, 21 (1993).
\bibitem{na94}  Y.V. Nazarov, Phys. Rev. Lett. {\bf 73}, 1420 (1994).
\bibitem{zai94} A.D. Zaikin, Physica B {\bf 203}, 255 (1994).
\bibitem{he93} F.W.J. Hekking and Y.V. Nazarov,
 Phys. Rev. Lett. {\bf 71}, 1625
 (1993).
\bibitem{he94} F.W.J. Hekking and Y.V. Nazarov, 
Phys. Rev. B {\bf 71}, 6847
 (1994).
\bibitem{be94}C.W.J. Beenakker, B.Rejaei, and J.A. Melsen,
 Phys. Rev. Lett. {\bf 72}, 2470 (1994).
\bibitem{ef80} K.B. Efetov, A.I. Larkin, and D.E. Khmelnitskii, 
Sov. Phys. JETP {\bf 52}, 568 (1980).
\bibitem{bcs} J. Bardeen, L.N. Cooper, and J.R. Schrieffer, 
Phys. Rev. B {\bf 108}, 1175 (57).
\bibitem{btk82} G.E. Blonder, M. Tinkham, and T.M. Klapwijk,
Phys. Rev. B. {\bf 25}, 4515 (1982).
\bibitem{la94} C.J. Lambert, Physica B {\bf 203}, 201 (1994),
 and references therein. 
\bibitem{crl95} N.R. Claughton, R. Raimondi,
 and C.J. Lambert , Phys. Rev. B {\bf 53},    (1996).
\bibitem{spi82} B.Z. Spivak and D.E. Khmel'nitskii,
 JETP Lett {\bf35}, 413 (1982).
\bibitem{al87} B.L. Al'tshuler and B.Z. Spivak, Sov. Phys. JETP
 {\bf 65}, 343 (1987).
\bibitem{la93} C.J. Lambert, 
J. Phys.: Condensed Matter {\bf 5}, 707 (1993).
\bibitem {hui93} V.C. Hui and C.J. Lambert,
 Europhys. Lett. {\bf 23}, 203
(1993).
\bibitem {za94} A.V. Zaitsev, Phys.lett.  A{\bf 194}, 315 (1994).
\bibitem{ta94} Y.Takane, J. Phys. Soc. Jpn. {\bf 63}, 2668 (1994).
\bibitem{asrl95} N.K. Allsopp, J. Sanchez-Cani\~zares, R. Raimondi
and C.J. Lambert  (unpublished).
\bibitem{na95} Y.V. Nazarov and T.H. Stoof, Phys. Rev. Lett. {\bf 76},
823 (1996).
\bibitem{ve94} P.G.N. de Vegvar, T.A. Fulton, W.H. Mallison, 
and R.E. Miller,
 Phys. Rev. Lett. {\bf 73}, 1416 (1994).
\bibitem{po94} H. Pothier, S. Gueron, D. Esteve, and M.H. Devoret,
Physica {\bf B203}, 226 (1994).
\bibitem {di95} A.Dimoulas et al., Phys. Rev. Lett. {\bf 74},
 602 (1995);
 B.J. van Wees et. al., Physica {\bf B203}, 285 (1994).
\bibitem{pe95} V. Petrashov et al., Phys. Rev. Lett. {\bf 74},
 5268 (1995).
\bibitem{scalapino95} For a recent review on the case of d-wave
pairing in high-$T_c$ superconductors, see D.J. Scalapino, 
Phys. Reports {\bf 250}, 329 (1995). 
\bibitem{note} We notice, however, that self-consistency in the presence
of d-wave pairing in hybrid superconducting structures has been
addressed explicitly by  Ch.Bruder,  Phys.Rev. B {\bf {41}},4017 (1990),
 and by  Y.S. Barash, A.V. Galaktionov, and A.D. Zaikin,
Phys. Rev. B {\bf 52}, 665 (1995).
See also the discussion in section V.
\bibitem{la91} C.J. Lambert, J. Phys.: Condensed Matter {\bf 3},
 6579 (1991).
\bibitem{lhr93} C.J. Lambert, V.C. Hui,
and S.J. Robinson, J.Phys.: Condens. Matter,
{\bf 5}, 4187 (1993).
\bibitem{sr92} M.Sigrist and T.M.Rice, J.Phys. Soc. Jpn. {\bf 61},
 4283 (1992).
\bibitem{wo93} D.A.D.Wollman et al.,  Phys. Rev. Lett. {\bf 71}, 2134 
(1993);
see also D.J.Van Harlingen  Rev. Mod. Phys. {\bf 67}, 515 (1995).
\bibitem{yu95} Y.Tanaka and S.Kashiwaya, Phys. Rev. Lett. {\bf 74},
 3451 (1995).
\bibitem{note2} According to the analysis by Y. S. Barash et al. (see
Ref.\cite{note}) when one of the crystal axes of the d-wave superconducting 
order parameter
is perpendicular to the interface plane, the value of the order
parameter in the proximity of the interface coincides with the equilibrium value
far from the interface itself.  

 
\end{references}
\end{document}